\documentclass[11pt]{article}
\usepackage[margin=1in]{geometry}
\usepackage{graphicx}
\usepackage[colorlinks=true,linkcolor=black]{hyperref}

\def\beq{\begin{equation}}
\def\eeq#1{\label{#1}\end{equation}}
\def\eeqn{\end{equation}}

\def\beqa{\begin{eqnarray}}
\def\eeqa#1{\label{#1}\end{eqnarray}}
\def\eeqan{\end{eqnarray}}

\let\bar=\overbar

\def\Dslash{\not{\hbox{\kern-4pt $D$}}}
\def\dslash{\not{\hbox{\kern-2pt $\del$}}}

\def\msb{{\bar{\ssstyle M \kern -1pt S}}}

\def\Title#1{\begin{center} {\Large {\bf #1} } \end{center}}
\def\Author#1{\begin{center} {\normalsize {\sc #1} } \end{center}}
\def\Institution#1{\begin{center} {\normalsize {\it #1} } \end{center}}
\def\Abstract#1{\noindent {\normalsize {\bf Abstract:} {\normalfont #1}}}
\def\Conference{\vspace{4mm}\begin{raggedright} {\normalsize {\it Talk presented at the 2019 Meeting of the Division of Particles and Fields of the American Physical Society (DPF2019), July 29--August 2, 2019, Northeastern University, Boston, C1907293.} } \end{raggedright}\vspace{4mm}}

\begin{document}

%
%

\Title{High-Pressure Gaseous Argon TPC for the DUNE Near Detector}

\Author{Kirsty Duffy, for the DUNE Collaboration}

\Institution{Fermi National Accelerator Laboratory}

\Abstract{The DUNE Near Detector design consists of multiple components, each designed to produce complementary constraints on flux and neutrino interaction systematic uncertainties for the oscillation analyses. One of these subdetectors is a magnetized high-pressure gaseous-argon TPC (HPgTPC), which will provide fine-grained tracking in a low-density detector, using the same target nucleus as the DUNE far detector. With its low momentum threshold for particle detection, the HPgTPC will be able to constrain one of the most crucial -- and least-well understood -- uncertainties for the oscillation analysis: nuclear effects in neutrino-argon interactions. These proceedings describe the current design, physics goals, and projected performance of the HPgTPC, as well as the ongoing R\&D work at Fermilab, in which a test-stand TPC is being built and will be operated at up to 10 atm pressure.}

\Conference

%
%

\section{Introduction}

\begin{figure}[htb]
\centering
\includegraphics[height=1.5in]{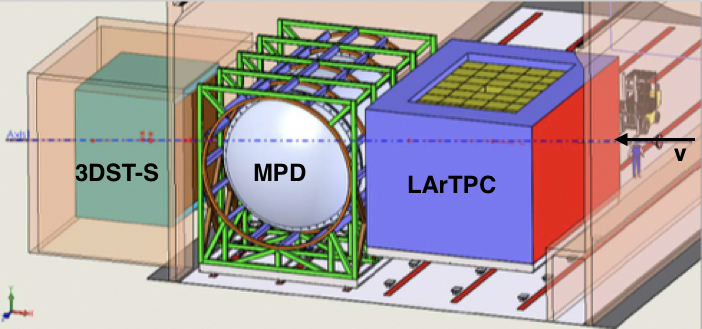}
\caption{Reference design for the DUNE near detector complex. From upstream (right) to downstream (left) according to the neutrino beam: the liquid argon time projection chamber (LArTPC), the multi-purpose detector (MPD), and the three-dimensional scintillating tracker-spectrometer (3DST-S). Figure from the DUNE collaboration, with annotation added by the author.}
\label{fig:DUNE_ND}
\end{figure}

The baseline design for the DUNE near detector--shown in figure~\ref{fig:DUNE_ND}--consists of multiple subdetector systems: a liquid argon time projection chamber (LArTPC), followed by a multi-purpose detector (MPD), and a three-dimensional scintillating tracker-spectrometer (3DST-S). The focus of these proceedings is the MPD, for which the design consists of a high-pressure gaseous argon TPC (HPgTPC), surrounded by an electromagnetic calorimeter (ECAL), in a 0.5T magnetic field provided by a superconducting Helmholtz-coil magnet.

The MPD has two important roles in the DUNE oscillation analysis. The first is to measure particles that exit the LArTPC, coming from neutrino interactions inside the liquid argon (LAr) fiducial volume. Because the DUNE far detectors will use LArTPC technology, one of the primary needs for the near detector is a high-statistics sample of neutrino interactions on argon. It is not feasible to contain all particles produced by a neutrino interaction within a LArTPC at the near detector -- the baseline design LArTPC is 7m wide x 3m high x 5m deep, and many particles produced in neutrino interactions, muons in particular, are expected to exit the detector. To account for this, the MPD, situated immediately downstream of the LArTPC, will measure the momentum of forward-going high-energy muons with high resolution using the particles' curvature in the magnetic field. The magnetic field will also allow the MPD to distinguish the sign of charged particles -- therefore distinguishing neutrino from antineutrino interactions, and measuring the wrong-sign component of the neutrino (or antineutrino) beam. This is particularly important for DUNE's electron antineutrino appearance measurement, for which a large component of the selected $\bar{\nu}_e$ sample comes from $\nu_{\mu}\rightarrow\nu_e$ oscillation of neutrinos in the antineutrino beam. Unlike the MPD, the LArTPC detectors that make up DUNE's near and far detectors will have no magnetic field, and will therefore not be able to distinguish between positively and negatively charged leptons.

\begin{figure}[htb]
\centering
\includegraphics[height=2.0in]{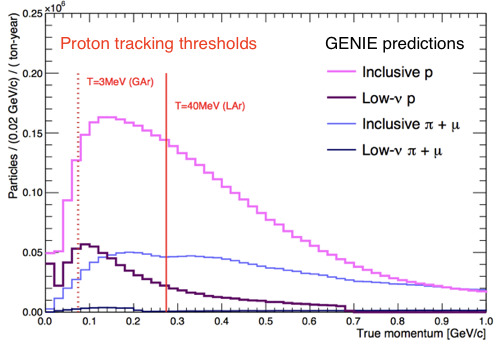}
\caption{Predicted distributions of protons, muons, and pions produced in some neutrino interactions, according to the GENIE generator. Proton reconstruction thresholds in liquid argon (LAr) and 10-atmosphere gaseous argon (GAr) are overlaid. Figure by L. Bellantoni.}
\label{fig:PThresh}
\end{figure}

The second important role of the MPD is to provide an independent sample of neutrino interactions on argon in the HPgTPC. Neutrino interactions in the argon gas will be on the same target and in the same beam as the near and far detector LArTPCs, but the HPgTPC allows for significantly lower reconstruction thresholds, as shown in figure~\ref{fig:PThresh}. 
The proton reconstruction threshold in liquid argon (LAr) is shown as 40~MeV of kinetic energy, informed by results from the MicroBooNE collaboration demonstrating automated reconstruction and identification of protons with kinetic energy 47~MeV and above~\cite{uBprotons}. The ArgoNeuT experiment was able to identify protons with 21~MeV kinetic energy~\cite{Acciarri:2014gev} using non-automated reconstruction, so there is reason to believe that the threshold in LAr can be pushed lower than the current MicroBooNE results. 
However, because gaseous argon is considerably lower density (even at high pressure) than liquid argon, particles with a given energy will travel much further in the HPgTPC than the LArTPC and will therefore be reconstructable at much lower energies. Existing reconstruction algorithms for the DUNE HPgTPC are able to reconstruct protons with as low as 3~MeV kinetic energy.
Lower thresholds for reconstructing particles will allow more accurate measurements of complex final states, leading to improved models of neutrino interactions in argon, and therefore reduced uncertainties for the oscillation measurement. 

An additional benefit of a sample of neutrino interactions in the HPgTPC is that this sample will have flat acceptance over the full angular range, mirroring the acceptance of the DUNE far detector. Neutrino interactions in the LArTPC can only be selected if a good measurement of the muon momentum is possible -- this requires that the muon is either low energy and contained within the detector, or very forward-going such that it exits the LArTPC and is measured in the MPD. By contrast, a determination of most particles' momentum is possible in the MPD by their curvature in the magnetic field, and therefore there is no angular requirement for interactions in the HPgTPC to be well reconstructed.

%

\section{A High-Pressure Gaseous Argon TPC as part of the DUNE Multi-Purpose Detector}

The focus of these proceedings is the high-pressure gaseous argon time projection chamber (HPgTPC), but it is worthwhile to describe the other components of the MPD to provide context, and because they will operate together as a single detector system.
Besides the HPgTPC, the MPD also includes an ECAL and a magnet. 

The current design for the MPD magnet uses Helmholtz coils to minimize material surrounding the HPgTPC (and, in particular, material between the LArTPC and the MPD). It will provide a 0.5~T magnetic field, which will enable distinction of positively from negatively charged particles and measurement of the momentum of exiting particles.

The ECAL, which surrounds the HPgTPC, is vital to measure the energy and direction of electromagnetic particles. Electrons and photons will produce showers in the ECAL, but in the HPgTPC those same particles will most likely leave tracks (in the case of electrons) or possibly no visible deposited energy (in the case of photons that don't convert in the gas). 
Therefore, ECAL information will allow the MPD to reconstruct photons, identify $\pi^0$s that may cause backgrounds to the $\nu_e$ selection in the LArTPCs, and identify external backgrounds from interactions outside the HPgTPC in which particles enter the detector. Studies are also ongoing to evaluate the ECAL's ability to detect neutrons produced by interactions in the argon gas. Importantly for the HPgTPC, the ECAL provides fast timing, with expected resolution on the order of 1~ns. This will allow us to determine when an interaction occurred and therefore reconstruct distances in the drift direction for the HPgTPC, and even to separate overlapping interactions. The current preliminary ECAL design uses tiles and strips of plastic scintillator. Scintillator tiles on the inner layers provide very good granularity and angular resolution. On the outer layers, where granularity is not so important, using strips of scintillator reduces the cost and number of readout channels in the detector.

The central detector of the MPD is the HPgTPC. The design is inspired by the ALICE TPC~\cite{Alme:2010ke}, which is currently being refurbished at CERN by upgrading its readout chambers. This means that the current readout chambers (consisting of around 600,000 readout channels) are no longer in use, and DUNE plans to repurpose them for the HPgTPC. The chambers were acquired in early/mid 2019 and are being prepared for shipping from CERN to Fermilab. 

Although the readout chambers will be reused, the current design is not simply a copy of the ALICE TPC -- a number of changes are planned. 
The ALICE design specifies a Ne-CO$_2$-N$_2$ gas mixture at one atm.~pressure (although ALICE has also run with other gas mixtures); 
DUNE plans to use an argon-based gas mixture (possibly 90:10 Ar:CH$_4$, although other mixtures are also under consideration) at a pressure of 10 atm. A 90:10 Ar:CH$_4$ gas mixture would give a 1~t fiducial mass for neutrino interactions in the HPgTPC, of which 97\% would occur on argon nuclei. The ALICE design has a central hole in the TPC to accommodate the inner vertex tracker and beam pipe; DUNE will use the same layout for the existing readout chambers, but new chambers will be designed and built to fill in this central region. As a new mechanical structure for the TPC will need to be built for DUNE, other design ideas from experiments such as sPHENIX and NA49 (for example, design of the field cage and calibration infrastructure) are under consideration.

The main reason to operate the TPC at high pressure is because it provides a larger-statistics sample of neutrino interactions in the gas. As previously mentioned, this is desirable because this will be an independent set of interactions on argon from those in the LArTPC, with different energy reconstruction and detector systematics, and lower detection thresholds. With a 1~t fiducial mass, we expect around 1.5$\times 10^6$ $\nu_{\mu}$CC interactions in the HPgTPC per year of neutrino-mode beam running. An additional, but secondary, benefit of higher pressure is that it allows for better particle identification by deposited energy per unit length both in the minimum-ionizing region and at the Bragg peak for particles that stop in the HPgTPC.

\begin{figure}[htb]
\centering
\includegraphics[height=3.0in]{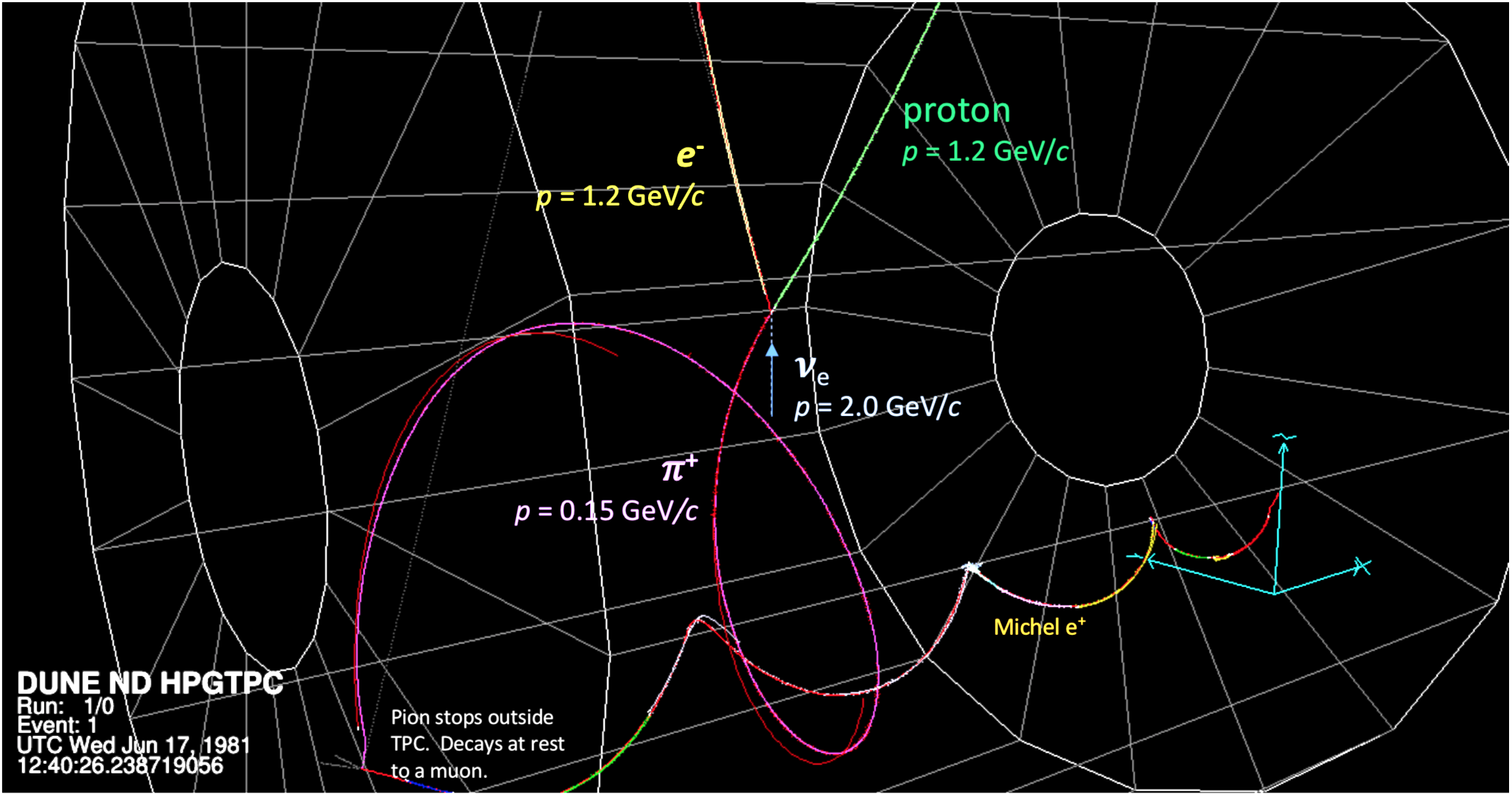}
\caption{Event display showing a simuluated and reconstructed $\nu_e$ interaction in the HPgTPC, in which an electron, $\pi^+$, and proton are detected. Figure by T. Junk.}
\label{fig:evtdisplay}
\end{figure}

The event display in figure~\ref{fig:evtdisplay} gives a demonstration of the data expected from the HPgTPC. This display shows a simulated and fully reconstructed $\nu_e$ charged-current interaction: $\nu_e + Ar \rightarrow e^- + \pi^+ + p + n$ (where the neutron does not leave a track in the HPgTPC). As well as demonstrating the finely-resolved data we expect of this detector, this event display also represents a huge amount of work by the collaboration to develop mature software and reconstruction for the HPgTPC.

%

\section{High-Pressure Gaseous Argon TPC Test Stands}

\begin{figure}[htb]
\centering
\includegraphics[height=2.5in]{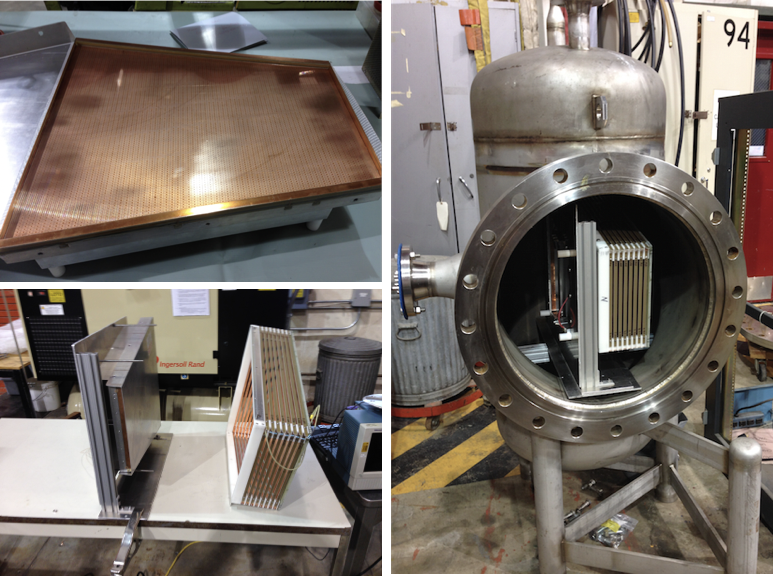}
\caption{Top left: the ALICE IROC. Bottom left: the IROC and field cage built to complete the TPC. Right: the IROC and field cage installed in the pressure vessel.}
\label{fig:GOAT}
\end{figure}

Tests are underway at two test stands to characterize and understand the ALICE readout chambers, and particularly their response at high pressure. Royal Holloway, University of London, has a test stand using one of ALICE's Outer Readout Chambers (OROC), which is 1~m long and 98~cm wide at its widest point. Here, we focus on the test stand at Fermilab that aims to characterize an Inner Read Out Chamber (IROC), as shown in figure~\ref{fig:GOAT}. The IROC is 40~cm long and 45~cm at its widest point. It has been installed in a pressure vessel rated for operation up to 10~atm pressure, with a purpose-built field cage to complete the TPC.

The first tests at Fermilab have been conducted at atmospheric pressure, using a $^{55}$Fe radioactive source installed on the back of the high-voltage cathode. A small hole was drilled in the cathode to enable X-rays from the source to enter the active volume. Signals are then read from groups of readout pads in two regions: a reference region towards the edge of the IROC, far away from the source, which should only see background signals; and an active region immediately below the source in the center of the IROC, which we expect to see a large signal due to the $^{55}$Fe X-rays. Using this setup, a measurement of the gas multiplication gain was made as a function of anode voltage at atmospheric pressure. This confirms the experimental setup of the Fermilab test stand by recreating measurements made by the ALICE collaboration, as well as providing a baseline for similar measurements at higher pressures.  

The next step for the Fermilab test stand will be to move to high pressure, increasing the pressure in the chamber from 1 to 10 atm.~in multiple steps, and repeating the gas multiplication gain measurement at each pressure setting. 
We know that the ALICE ROCs work at atmospheric pressure, and electrostatic calculations predict they should also work well at higher pressure. Still, demonstrating successful operation at high pressure with the as-built chambers will be a significant proof of principle for the DUNE near detector design. Fermilab clearance for high-pressure operation was granted at the end of July 2019, and tests at higher pressures are currently ongoing.

%

\section{Summary}

These proceedings discuss the high-pressure gaseous-argon TPC that will form part of the DUNE multi-purpose detector, along with an ECAL and a magnet system providing a 0.5~T magnetic field. The MPD is a vital component of the DUNE near detector, which provides high-resolution measurements of particles exiting the LArTPC as well as an independent sample of interactions on the same target in the same beam, with different detector effects and much lower thresholds. 

The baseline HPgTPC design is to reuse readout chambers from the ALICE TPC, with a full TPC design inspired by ALICE and incorporating additional ideas from other running gaseous TPC experiments. Two test stand efforts are underway to characterize the ALICE readout chambers and prove the feasibility of high-pressure operation.

%
\clearpage
\section*{Acknowledgements}

These proceedings show work by many members of the DUNE collaboration, and in particular the MPD and GOAT working groups. Particular thanks are due to J. Raaf, T. Junk, T. Mohayai, L. Bellantoni, E. Brianne, A. Bross, and all staff at the PAB in Fermilab where the GOAT test stand is based.



\end{document}